\documentclass[aps,prb,groupedaddress,twocolumn,showpacs]{revtex4}
\usepackage{graphicx}

\begin{document}

\title{ Resistive transition in frustrated Josephson-junction arrays
on a honeycomb lattice}

\author{Enzo Granato}

\address{Laborat\'orio Associado de Sensores e Materiais,
Instituto Nacional de Pesquisas Espaciais, 12227-010 S\~ao Jos\'e
dos Campos, SP Brazil}

\begin{abstract}
We use driven Monte Carlo dynamics to study the resistive behavior
of superconducting Josephson junction arrays on a honeycomb lattice
in a magnetic field corresponding to $f$ flux quantum per plaquette.
While for $f=1/3$ the onset of zero resistance is found at nonzero
temperature, for $f=1/2$ the results are consistent with a
transition scenario where the critical temperature vanishes and the
linear resistivity shows thermally activated behavior. We determine
the thermal critical exponent of the zero-temperature transition for
$f=1/2$, from a dynamic scaling analysis of the nonlinear
resistivity. The resistive behavior agrees with recent results
obtained for the phase-coherence transition from correlation length
calculations  and with experimental observations on ultra-thin
superconducting films with a triangular pattern of nanoholes.
\end{abstract}

\pacs{74.81.Fa, 74.25.Uv, 75.10.Nr}

\maketitle

\section{Introduction}

Josephson-junction (JJ) arrays have remarkable properties in a
magnetic field, which are strongly dependent on the geometry of the
structure. In addition to being realized as two-dimensional arrays
of weakly coupled superconducting grains \cite{tinkham,zant,baek},
they provide important models for superconducting wire networks
\cite{giroud,yu,ling,xiao} and other inhomogeneous superconduting
systems, when phase fluctuations of the superconducting order
parameter play a major role \cite{emery}. An idealized JJ array is
equivalent to the frustrated XY model \cite{teitel}, where
frustration can be tuned by the applied external magnetic field. The
frustration parameter $f$, corresponding to the number of flux
quantum per plaquette of the array, sets the average density of
vortices in the lattice of pinning sites formed by the plaquette
centers. Depending on the topology of the lattice of pinning sites
and the value of $f$, a commensurate vortex lattice is favored in
the ground state, allowing for a phase-coherence transition at
finite temperature. In this case, the equilibrium phase transitions
and resistive behavior of the superconducting  array  are reasonably
well understood for simple low-order commensurate phases such as
$f=1/2$ on a square array \cite{teitel} and $f=1/3$ on a honeycomb
array \cite{shih85}. The magnetoresistance for a square JJ array,
for example, oscillates with the applied magnetic field
\cite{tinkham,zant,ling}, displaying minima at integer values of $f$
and secondary minima at $f=1/2$ for decreasing temperatures,
corresponding to resistive transitions at different temperatures
\cite{teitel}. The onset of zero-resistance for decreasing
temperatures marks the phase-coherence transition in the JJ array,
which for integer $f$ is expected to be in the Koterlitz-Thouless
(KT) universality class. Dynamical transitions under an external
driving current have also been studied for $f=1/2$ on a square
lattice, leading to interesting nonequilibrium phase diagrams
\cite{driven}. However, when the vortex lattice is incommensurate
with the pinning sites, as for irrational $f$ on a square JJ array
\cite{baek,yu,ling,halsey,teitelf,park,eg07,eg08} or $f=1/2$ on a
honeycomb JJ array \cite{xiao,shih85,reid,korshu,eg12}, the possible
phase transitions are much less understood, showing some features of
a vortex glass without disorder and dynamical freezing at low
temperatures. In particular, a JJ array on a honeycomb lattice with
$f=1/2$, should display interesting resistive behavior. As a model
of phase fluctuations, it should be relevant to ultra-thin
superconducting films with a periodic pattern of nanoholes
\cite{valles,baturina}, which can be regarded as a lattice of
pinning centers. While for a square lattice of nanoholes, the
magnetoresistance oscillates with the applied field,  displaying
secondary minima at $f=1/2$ as for a square JJ array
\cite{baturina}, for a triangular lattice \cite{valles} it shows
only minima at integer flux quantum per lattice unit cell.

In early Monte Carlo (MC) simulations of the fully frustrated XY
model on a honeycomb lattice \cite{shih85}, a phase-coherence
transition at a nonzero temperature in the KT universality class was
suggested and therefore a resistive transition would be expected for
a JJ array in the same lattice with $f=1/2$. On the other hand, a
different calculation \cite{reid} suggested a spin-glass like
transition.
%approximately at the same temperature, which could lead
%to a resistive transition in a different universality class.
It was also suggested \cite{leeteitel} that only a crossover region
rather than an equilibrium phase transition should occur at any
nonzero temperature.  Recently \cite{korshu}, it was argued that
vortex-ordered phases could be possible at nonzero temperatures but
for very large systems, beyond the ones currently studied
numerically or even experimentally. However, the question of the
resistive transition was not investigated. In a recent MC study of
phase coherence in the fully frustrated XY model a zero-temperature
transition scenario \cite{eg12} was proposed, where $T_c=0$ but the
divergent correlation length, $\xi \propto T^{-\nu}$, should lead to
measurable effects at finite temperatures in the linear and
nonlinear resistivity, determined by the  thermal critical exponent
$\nu$. So far, a direct calculation of the resistive behavior of JJ
arrays on a honeycomb lattice and comparison to experiments have not
been presented.

In this work, we present results for the resistive behavior obtained
by driven Monte Carlo dynamics. While for $f=1/3$ a resistive
transition is found at nonzero temperature, for $f=1/2$ the results
are consistent with a transition scenario where the critical
temperature vanishes and the linear resistivity shows thermally
activated behavior. We determine the thermal critical exponent $\nu$
of the zero-temperature transition for $f=1/2$, from a dynamic
scaling analysis of the nonlinear resistivity. Its value is in fair
agreement with recent calculations for the frustrated XY model from
finite-size correlation length scaling \cite{eg12}. A dynamical
freezing at lower temperatures is also identified from deviations of
the fluctuation-dissipation relation between linear resistivity and
voltage autocorrelations. The resistive behavior is consistent with
some experimental observations in ultra-thin superconducting films
with a triangular lattice of nanoholes \cite{valles}, taking into
account the effects of weak Josephson-coupling disorder.

\begin{figure}
\includegraphics[bb= 0cm -0cm  7cm   5.5cm, width=7.5cm]{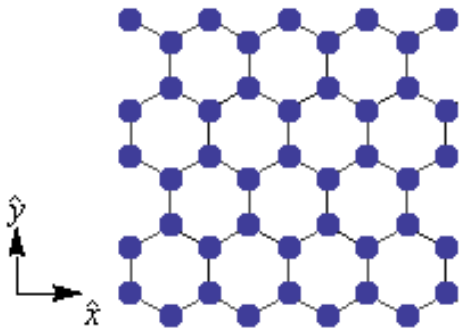}
\caption{ JJ array on a honeycomb lattice. Filled circles represent
superconducting grains and the lines the Josephson junctions between
them. } \label{honey}
\end{figure}

\section{Model and Driven Monte Carlo Simulation}

We consider a JJ array in a uniform transverse magnetic field
described by the Hamiltonian
\begin{equation}
H=-\sum_{<ij>}J_{ij}\cos(\theta_i -\theta_j-A_{ij}) -J
\sum_i(\theta_i-\theta_{i+\hat x}) , \label{model}
\end{equation}
where $\theta_i$ is the phase of the local superconducting order
parameter of the grains located on  the sites of a two-dimensional
honeycomb lattice with lattice spacing $a$, as illustrated in   Fig.
\ref{honey}. The first term is the contribution from the
Josephson-coupling energy between nearest neighbor grains. For
uniform coupling we set $J_{ij}=J_o$, a constant independent of the
magnetic field.  The line integral of the vector potential $A_{ij}$
due to the external field $\vec B = \nabla \times \vec A$ is
constrained to $\sum_{ij}A_{ij} = 2 \pi f$ around each hexagonal
plaquette, where $f$ is the number of flux quantum $\phi_o=hc/2e$
per plaquette. This model is periodic in $f$ with period $f = 1$. In
the calculations we choose a gauge where $A_{ij} = 2 \pi f n_i/2$ on
the (tilted) bonds along the horizontal rows numbered by the integer $n_i$
and $A_{ij}=0$ on the vertical bonds of the lattice. The second term
in Eq. (\ref{model}) represents the effects of an external driving
current density $(2 e/\hbar) J$ applied in the $\hat x$ (horizontal)
direction, coupling to the phase difference,
$\theta_i-\theta_{i+\hat x}$, between nearest neighbors sites in
this direction.  When $J\ne 0$, the total energy is
unbounded and the system is out of equilibrium. The lower-energy
minima occur at phase differences $\theta_i -\theta_{i+\hat x}$ which
increases with time $t$, leading to a net phase slippage rate
proportional to $< d(\theta_i -\theta_{i+\hat x})/dt>$, corresponding to
the voltage $V_{i,i+\hat x}$. For convenience, we use units where $2 e
/\hbar=1 $, $J_o=1$ and $a=1$.

%\begin{figure}
%\includegraphics[bb= -0.5cm -0cm  6cm   4.5cm, width=7.5cm]{honeyplot.eps}
%\caption{ Josephson-junction array on a honeycomb lattice. Filled
%circles represent superconducting grains and the lines the Josephson
%junctions between them.}\label{honey}
%\end{figure}

\begin{figure}
\includegraphics[bb= 1cm -0cm  9cm   6.5cm, width=7.5cm]{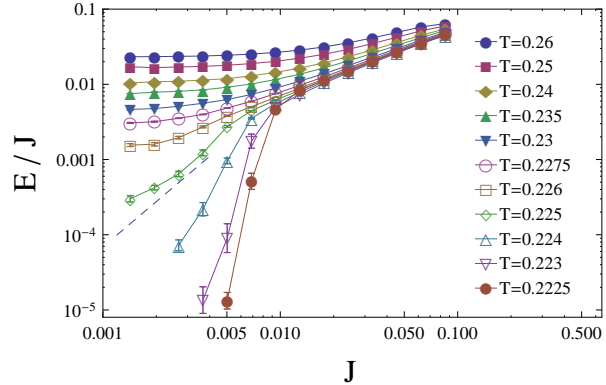}
\caption{ Nonlinear resistivity $E/J$ as a function of current
density $J$ at different temperatures $T$, for $f=1/3$. System size $L=60$.
The dashed line indicates  power-law behavior $ E \propto J^3$ .
}\label{nlf13}
\end{figure}

To study the current-voltage behavior, we use a driven MC dynamics
method \cite{eg04}. The time dependence is obtained by identifying
the MC time as the real time $t$ and we set the unit of time $dt=1$,
corresponding to a complete MC pass through the lattice. For
convenience, the honeycomb lattice is defined on a rectangular
geometry (Fig. \ref{honey}), with linear size given by a
dimensionless length $L$. In terms of $L$, the linear size in the
$\hat x$ and $\hat y$ directions can be written as $L_x=L \sqrt{3}
a$ and  $L_y= \frac{3}{2} a$, respectively. This corresponds to $2
L$ junctions along the horizontal rows. The usual periodic boundary
conditions are used in the $\hat y$-direction and periodic
(fluctuating twist) boundary conditions \cite{saslow} in the $\hat
x$ direction. The twist boundary condition adds a new dynamical
variables $u_x$, corresponding to a uniform phase twist between
nearest-neighbor sites along the $\hat x$ direction. A MC step
consists of an attempt to change the local phases $\theta_i$ and the
phase twist $u_x$, using the Metropolis algorithm. If the change in
energy is $\Delta H$, the trial move is accepted with probability $
min\{1,\exp(-\Delta H/kT)\}$. The external current density $J$ in
Eq. (\ref{model}) biases these changes, leading to a net voltage
(phase slippage rate) across the system in the $\hat x$-direction
given by
\begin{equation}
 V = 2 L \frac{d}{dt}  u_x  ,
\end{equation}
in arbitrary units.  Compared to the usual Langevin dynamics
\cite{eg07}, this MC method allows to  access  much longer time
scales, which is required to  obtain reliable data at lower
temperatures and current densities. We have determined the electric
field $E=V/(2L)$ and nonlinear resistivity $\rho = E/J$ as a
function of the driving current density $J$, in the $\hat x$
direction, for different temperatures $T$ and different system sizes
$L$. We used typically $ 5 \times 10^6$ MC steps to reach the
nonequilibrium steady state and equal time steps to perform time
averages, with additional averages over $6-12$ independent runs. In
a MC step, the maximum changes in the local phases $\theta_i$ and
the phase twist $u_x$ were fixed to $\pm \pi$ and $\pm \pi/(2 L )$,
respectively.

The linear resistivity, $\rho_L=\lim_{J->0} E/J$, can be determined
from the nonlinear behavior $\rho(J)$, obtained from the driven MC
simulations, by extrapolating the numerical results to vanishing
currents. It can also be obtained, independently, from equilibrium
voltage fluctuations and therefore can be calculated in absence of
an imposing driving current ($J=0$). From Kubo formula, the linear
resistance is given in terms of the equilibrium voltage
autocorrelation as
\begin{equation}
R_L=\frac{1}{2T} \int d t \langle V(t) V(0) \rangle . \label{kubo}
\end{equation}
Since the total voltage $V$ is related to the phase difference
across the system $\Delta \theta(t)$ by $V= d\Delta \theta(t)/dt$,
we find more convenient to determine $R_L$ from the long-time
equilibrium fluctuations \cite{eg98} of $\Delta \theta(t)$ as
\begin{equation}
R_L=\frac{1}{2Tt} \langle (\Delta \theta(t)-\Delta
\theta(0))^2\rangle  , \label{rho}
\end{equation}
which is valid for sufficiently long times $t$.
%MC simulations for
%the linear resistivity obtained from Eq. (\ref{rho}) were performed
%using typically $2 \times 10^3$ time averages for $4 \times 10^6$ MC
%steps.

\begin{figure}
\includegraphics[bb= 1cm -0cm  9cm   6.5cm, width=7.5cm]{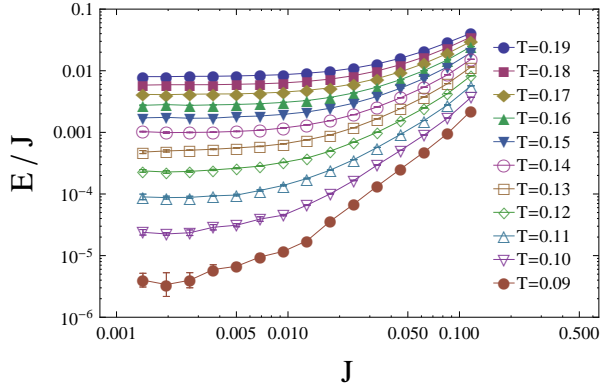}
\caption{ Nonlinear resistivity $E/J$ as a function of current
density $J$ at different temperatures $T$, for $f=1/2$. System size $L=60$. }
\label{nlf12}
\end{figure}

\section{Results and Discussion}

First, we consider the resistive behavior when $f=1/3$. For this
value of the frustration, it is known that a hexagonal vortex
lattice commensurate with the honeycomb lattice is the ground state
\cite{shih85}, and therefore a resistive transition would be
expected at a temperature smaller or equal the vortex lattice
melting. Fig. \ref{nlf13} shows the nonlinear resistivity $E/J$ as a
function of temperature, for the largest system size $L=60$, where
finite-size effects are small. For decreasing current densities $J$,
the nonlinear resistivity $E/J$ tends to a finite value at high
temperatures, corresponding to the linear resistivity $\rho_L$, but
extrapolates to very low values at lower temperatures. This behavior
is consistent with a resistive transition occurring at a critical
temperature in the range $T_c(f=1/3)=0.224-0.225$. In fact, it is
slightly smaller than the vortex lattice melting transition
estimated from recent equilibrium MC simulations of the frustrated
XY model on a honeycomb lattice \cite{eg12}, $T_m = 0.226(1)$. At
the resistive transition, a power-law relation  $ E \propto J^{z+1}$
is expected at sufficiently small currents from the scaling theory
\cite{fisher}, where $z$ is the dynamical critical exponent. For the
usual KT transition it is known \cite{tinkham,fisher} that $z=2$. In
the present case, as shown by the dashed line in Fig. \ref{nlf13}, a
power-law  separating the $T>T_c$ from $T < T_c$ behavior at small
currents is compatible with $z=2$. However, further work taking into
account finite-size effects is required to investigate the critical
behavior in detail. In any case, the above results show clear
evidence of a resistive transition at finite temperature for
$f=1/3$.

In contrast to the resistive behavior in Fig. \ref{nlf13}, when
$f=1/2$ the nonlinear resistivity $E/J$ tends to a finite value for
decreasing currents even at low temperatures as shown in Fig.
\ref{nlf12}. Although we can not exclude a transition at much lower
temperatures, where reliable data could not be obtained as discussed
below, this behavior is consistent with a resistive transition
occurring only at zero temperature. Recent equilibrium MC
simulations suggested such zero-temperature transition scenario
\cite{eg12}, where $T_c=0$ for the phase-coherence transition but
the finite correlation length for $T>0$ leads to measurable effects
in the nonlinear resistivity. In fact, the behavior in Fig.
\ref{nlf12} has the main features expected for a zero-temperature
resistive transition. The linear resistivity $\rho_L$, corresponding
to zero current limit of $E/J$, decreases rapidly with decreasing
temperature and for increasing $J$,  $E/J$ cross over to a nonlinear
behavior at a characteristic current density $J_{nl}$, which also
decreases with decreasing temperature.

\begin{figure}
\includegraphics[bb= 1cm -0cm  9cm   6.5cm, width=7.5cm]{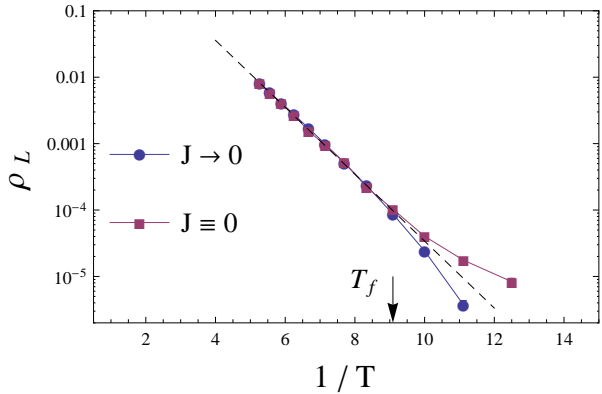}
\caption{ Temperature dependence of the linear resistivity $\rho_L$
for $f=1/2$ obtained from nonlinear resistivity ($J\rightarrow 0$)
and from voltage fluctuations ($J \equiv 0$). System size $L=60$.
The separation of the curves gives an estimate of the dynamical freezing temperature
$T_f$. The dashed line is an Arrhenius fit for $T > T_f$.
}\label{lrf12}
\end{figure}

To verify in which temperature range the values approached at low
currents in Fig. \ref{nlf12} correspond indeed to the linear
resistivity $\rho_L$, we show in Fig. \ref{lrf12} the temperature
dependence of $\rho_L$ obtained from the nonlinear resistivity as
$\rho_L=\lim_{J->0} E/J$ and, without current bias, from Eq.
(\ref{kubo}).  These values obtained from nonequilibrium and
equilibrium calculations agree with each other above a temperature
$T_f\sim 0.11$ and deviate significantly at lower temperatures.
Since this agreement is only expected when the voltage
autocorrelation in Eq. (\ref{kubo}) is obtained in true equilibrium,
one can regard $T_f$ as a signature of a dynamical freezing
transition, below which equilibrium is not achieved due to very
large relaxation time. Interestingly, a dynamical freezing
transition near the same temperature was also identified in recent
equilibrium MC simulations by other methods and different dynamics
\cite{eg12}. The apparent KT transition \cite{shih85} and spin-glass
transition \cite{reid} observed in earlier MC simulations could be
attributed to slow dynamics effects of such dynamical freezing.

\begin{figure}
\includegraphics[bb= 1cm -0cm  9cm   6.5cm, width=7.5cm]{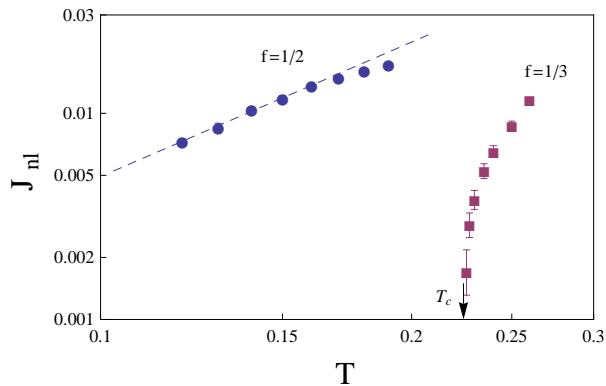}
\caption{ Temperature dependence  of the crossover current $J_{nl}$
for $f=1/2$ and $f=1/3$. System size $L=60$.
The dashed line is a power-law fit to $J_{nl}\propto  T^{1+\nu }$, giving the estimate $\nu=1.17(14)$.
The arrow indicates the estimated critical temperature for $f=1/3$. } \label{jnl}
\end{figure}

The straight-line behavior of $\rho_L(T)$ for $ T > T_f$ in the
log-linear plot of Fig. \ref{lrf12} indicates an activated Arrhenius
behavior, where the linear resistivity decreases exponentially with
the inverse of temperature with a temperature-independent energy
barrier, estimated as  $E_b = 1.16(4) J_o$. If such behavior
extrapolates to lower temperatures, it suggests that the linear
resistivity can be very small  but nevertheless remains finite for
decreasing temperatures  and therefore there is no resistive
transition at finite temperatures. However, as will be described
below, the system behaves as if a resistive transition occurs at
zero temperature, corresponding to a phase-coherence transition
where the critical temperature vanishes, $T_c=0$.

\begin{figure}
\includegraphics[bb= 1cm -0cm  9cm   6.5cm, width=7.5cm]{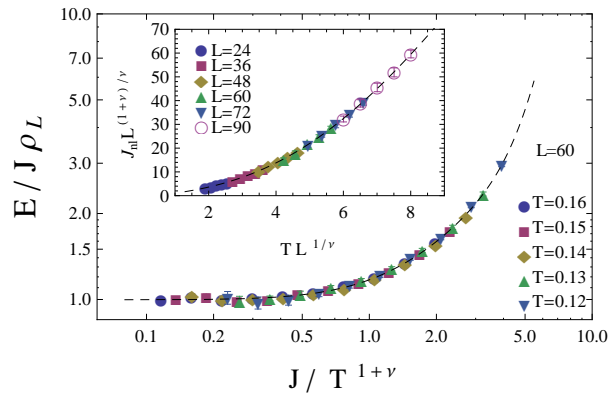}
\caption{ Scaling plot of the nonlinear resistivity $E/J$ from Fig.
\ref{nlf12}, system size $L=60$, for different temperatures above $T_f$ and
current range $J < 0.03$, giving the estimate $\nu =1.40(9)$. Inset: Finite-size
scaling plot of the crossover current $J_{nl}$ for different
temperatures $T$ and system sizes $L$, giving the estimate $\nu =1.15(9)$.
The dashed line is the fit used in the data collapse procedure. }
\label{scalnl}
\end{figure}

A detailed scaling theory \cite{fisher} of the resistive transition
with $T_c=0$ has been described in the context of the current-voltage
characteristics of vortex-glass models \cite{eg98,fisher,wengel} of disordered
two-dimensional superconductors  but the arguments should also apply
to the present case. The basic assumption is the existence of a
second-order phase transition. The correlation length $\xi$ is
finite for $T>0$ but it increases with decreasing temperature as
$\xi \propto T^{-\nu}$, with $\nu$ a critical exponent. The
divergent correlation length and relaxation time $\tau$ near the
transition determine both the linear an nonlinear resistivity
behavior leading to current-voltage scaling sufficiently close to
the critical temperature and sufficiently small driving current. If
the data satisfy such scaling behavior for different driving
currents and temperatures, the critical temperature and critical
exponents of the underlying equilibrium transition at $J=0$ can then
be determined from the best data collapse.  The dimensionless ratio
$E/J\rho_L$ should satisfy the scaling form \cite{fisher}
\begin{equation}
\frac{E}{J \rho_{L}}=g(\frac{J}{T^{1+\nu} }) \label{scalt0}
\end{equation}
where $g$ is a scaling function with $g(0) =1$. A crossover from
linear behavior, when $g(x) \sim 1 $, to nonlinear behavior, when
$g(x) >>1$, occurs when $x \sim 1$ which leads to a crossover
current density at which nonlinear behavior sets in, decreasing with
temperatures as a power law, $J_{nl}\propto T/\xi \propto T^{1+\nu
}$. The scaling form in Eq. (\ref{scalt0}) contains a single
critical exponent $\nu$ and does not depend on the particular form
assumed for the divergence of the relaxation time $\tau$. However,
for sufficiently low temperatures, the relaxation process is
expected to be thermally activated \cite{fisher} with $\tau \propto
\exp(E_b/kT)$. This corresponds formally to a dynamic exponent $z
\rightarrow \infty$, if power-law behavior is assumed for the
relaxation time $\tau \propto \xi^z$. The linear resistivity should
scale as \cite{fisher} $\rho_L \propto 1/\tau$ and therefore it is
also expected to have an activated behavior, $\rho_L \propto
\exp(-E_b/kT)$.  In general, the energy barrier $E_b$ also scales
with the correlation length as $E_b \propto \xi^\psi $, which leads
to a temperature-dependent barrier $E_b \propto T^{-\psi \nu }$. A
pure Arrhenius behavior corresponds to $\psi = 0$.

\begin{figure}
\includegraphics[bb= 1cm -0cm  9cm   6.5cm, width=7.5cm]{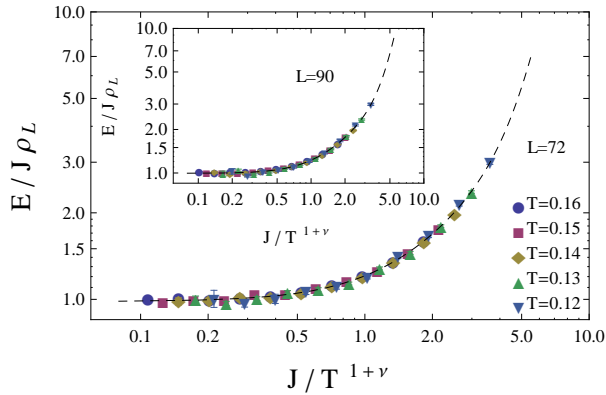}
\caption{ Scaling plot of the nonlinear resistivity $E/J$ as in Fig.
\ref{scalnl}, for a larger system size $L=72$, giving the
estimate  $\nu =1.36(8)$.
Inset: Same scaling plot for $L=90$, giving the estimate $\nu =1.33(8)$. }
\label{scalnlL}
\end{figure}

The behavior of the nonlinear and linear resistivity in Figs
\ref{nlf12} and \ref{lrf12} above the dynamical freezing temperature
$T_f$ are quite consistent with the predictions from the scaling
theory. Fig. \ref{jnl} shows the temperature dependence of the
crossover current $J_{nl}$, defined as the value of $J$ where $E/J
\rho_L$ starts to deviate from a fixed value, chosen to be $c=1.2$.
For the lowest temperature range above $T_f$, the linear behavior in
the log-log plot is consistent with the expected power-law
$J_{nl}\propto  T^{1+\nu }$ for a zero-temperature transition. From
the  power-law fit we obtain a first estimate of the exponent $\nu =
1.17(14)$. In contrast, for $f=1/3$,  the behavior in the lowest
temperature range does not allow a similar power-law fit; $J_{nl}$
curves down for decreasing temperatures and extrapolates to zero at
a finite temperature, consistent with a resistive transition at a
nonzero critical temperature found for $f=1/3$. The nonlinear
resistivity data also satisfies the scaling form for different
driving currents and temperatures. Fig. \ref{scalnl} shows a scaling
plot of the nonlinear resistivity above $T_f$ according to Eq.
(\ref{scalt0}), for a large system size $L=60$, where the
finite-size dependence is small. The best data collapse provides an
estimate of the critical exponent $\nu = 1.40(9)$. The data collapse
is achieved quantitatively by means of a least-squares fit method
\cite{eg12,night}, varying the parameter $\nu$. The scaling function
$g(x)$ is approximated by a Taylor series expansion for small $x$,
truncated beyond 4th order, which is used to fit the data and
provide the least-square residuals. The error estimate here
corresponds to the statistical error from the least-squares method
and do not include systematic effects. To check for systematic
errors from finite-size effects, which were assumed negligible in
the scaling form of Eq. (\ref{scalt0}), the same data collapse
procedure was repeated  for larger system sizes, as shown in Fig.
\ref{scalnlL}. The results of these estimates, $\nu = 1.36(8)$ for
$L=72$ and $1.33(8)$ for $L=90$, agree within the statistical errors
but indicate that the central estimate of $\nu$ decreases slowly
with system size. The nonlinear resistivity should also satisfy the
expected finite-size behavior in smaller system sizes when the
correlation length $\xi$ approaches the system size $L$. According
to finite-size scaling, the scaling function in Eq. (\ref{scalt0}),
should also depend on the dimensionless ratio $L/\xi$ and so, to
account for finite-size effects, the nonlinear resistivity should
satisfy the scaling form
\begin{equation}
\frac{E}{J \rho_{L}}=\bar{g}(\frac{J}{T^{1+\nu} },L^{1/\nu}T).
\label{scaltL}
\end{equation}
The scaling analysis of the whole nonlinear resistivity data is
rather complicated in this case since the scaling function depends
on two variables. To simplify the analysis \cite{wengel} we first
estimate the temperature and finite-size behavior of the crossover
current density $J_{nl}$ where nonlinear behavior sets in, as the
value of $J$ where $E/J \rho_L = c$, a constant. Then, from Eq.
(\ref{scaltL}), the finite-size behavior of $J_{nl}$ can be
expressed in the scaling form
\begin{equation}
J_{nl} L^{(1+\nu)/\nu}=\bar{\bar{g}}(L^{1/\nu}T).
\label{scaltnl}
\end{equation}
The best data collapse according to the scaling in Eq.
(\ref{scaltnl}) provides an independent estimate of the critical
exponent $\nu$. The inset in Fig. \ref{scalnl}  shows that indeed
the values of $J_{nl}$ for different system sizes and temperatures
satisfy this scaling form with $\nu = 1.15(9)$. To check for
systematic errors due to corrections to finite-size scaling, the
data collapse was repeated dropping the smaller system sizes.
Dropping system size $L=24$ gives $\nu =1.17(7)$ and $L=24$ to
$L=36$ gives $\nu =1.16(7)$. Since the resulting changes are small
compared with the errorbars, systematic errors of this kind are not
significant for this range of system sizes. The two independent
estimates of $\nu$ obtained above, 1.33(8) from the largest system
size and 1.15(9) from finite-size scaling, are not compatible within
the estimated errors. However, the former value could still be
affected by finite-size effects. The latter value should be more
accurate since it is based on finite-size scaling. This value is in
reasonable agreement, within the estimated errors, with the critical
exponent for the zero-temperature phase-coherence transition,
$\nu_{ph}=1.29(15)$, of the frustrated XY model obtained recently by
correlation length calculations using equilibrium MC simulations
\cite{eg12}. The Arrhenius behavior for the linear resistivity
$\rho_L$ in Fig. \ref{lrf12} is also consistent with the exponential
divergence of the relaxation time $\tau$  found in the equilibrium
MC simulations.

\begin{figure}
\includegraphics[bb= 1cm -0cm  8cm   5.5cm, width=7.5cm]{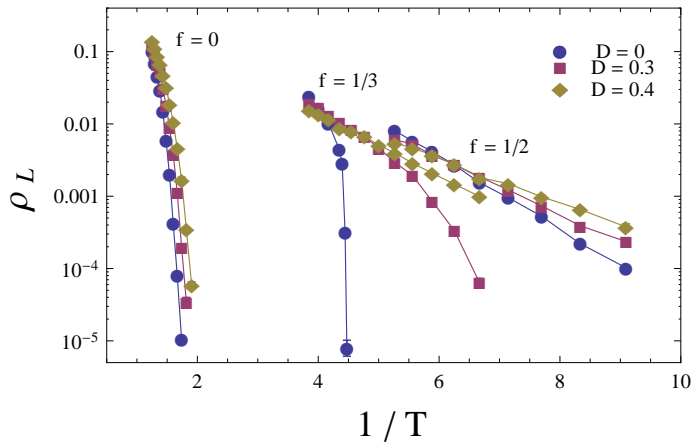}
\caption{ Temperature dependence of the linear resistivity $\rho_L$
for different frustration parameters $f$ and coupling disorder strengths $D$.
System size $L=60$.
}\label{lrdisor}
\end{figure}

Some experimental observations on ultra-thin superconducting films
with a triangular pattern of nanoholes \cite{valles} are consistent
with the zero-temperature resistive transition for $f=1/2$. In the
regime where phase fluctuations of the superconducting order
parameter are more important than amplitude fluctuations
\cite{emery,giroud}, this system can be described by an array of
superconducting "grains" coupled by  Josephson junctions in a
suitable geometry. The simplest model consists of a
Josephson-junction array on a honeycomb lattice, with the triangular
lattice of nanoholes corresponding to the lattice of pinning sites
(plaquette centers in Fig. \ref{honey}) and the number of flux
quantum per unit cell of the nanohole lattice corresponding to the
frustration parameter $f$ of the array. In fact, the measured
resistance of samples which are superconducting at low temperatures
and low magnetic fields, oscillates as a function of the magnetic
field, displaying minima at integer values of $f$ but no secondary
minima at$f=1/2$, as expected from the present results for the
honeycomb JJ array. Moreover, for $f=1/2$, the temperature
dependence of the resistance shows the expected Arrhenius behavior,
consistent with a vanishing critical temperature. However, the
measured magnetoresistance does not display minima at $f=1/3$, which
would be expected from the above calculations for temperatures near
$T_c(f=1/3)$. Although the available temperatures in the experiments
may not be sufficiently small to observe this feature, it could also
be the effect of quenched disorder in the Josephson couplings. In
fact, it was recently suggested that inhomogeneities in the film
thickness could lead to significant variations in the weak links
between superconducting islands \cite{valles}.
%Weak disorder in the
%Josephson coupling $J_{ij}$ should have a significant qualitative
%effect in the commensurate vortex phase at $f=1/3$ but minor effects
%at $f=0$, where the transition is in the KT universality class, and
%at $f=1/2$, where the critical temperature vanishes.

We have performed additional calculations to verify the qualitative
effect of weak disorder of the Josephson couplings on the
magnetoresistive behavior. We consider a simple random-coupling
model, where $J_{ij}$ in Eq. (\ref{model}) is defined as $J_{ij}=
J_o(1 \pm D) $, with equal probability, and  disorder strength
parameter $D$. The JJ array is still assumed to be on a perfect
honeycomb lattice. The resistivity as a function of temperature was
calculated by averaging over different realizations of the disorder.
Fig. \ref{lrdisor} compares the temperature dependence of $\rho_L$
obtained without current bias, from Eq. (\ref{kubo}), for $f=0,
f=1/3$ and  $f=1/2$,  and different disorder strengths $D$. While
the behavior characteristic of a finite-temperature transition for
$f=0$ and zero-temperature transition for $f=1/2$ remain for
increasing disorder, the resistive behavior for $f=1/3$ changes to
an Arrhenius form  above a disorder strength $D \sim 0.35$. In this
case, the magnetoresistance should only display minima at integer
values of $f$, as observed experimentally \cite{valles}, which in
turn suggests that coupling disorder should also play an important
role in modeling other phase coherence properties of this system.
When comparing the disorder strength $D$ in the  model with the
thickness variations in the experimental system, geometrical
disorder in the JJ array due to spatial irregularities of the
system, should also be taken into account. Weak positional disorder
of the grains, for example, has significant effects both on
phase-coherence and vortex order at nonzero values of frustration
\cite{gk86,benz88,korsh96,gupta99}, even when the Josephson coupling
is uniform ($D=0$). One then would expect that the combined effect
of geometrical and Josephson coupling disorder in the model will
result in an  Arrhenius behavior for $f=1/3$ occurring at much lower
values of $D$. These interesting effects and a more quantitative
comparison of such disorder model with the experimental system
deserves further work.

\section{Conclusions}

We have investigated  the resistive behavior of Josephson junction
arrays on a honeycomb lattice using driven MC dynamics, focusing
mainly on the $f=1/2$ frustration and its relation to experiments on
ultra-thin superconducting films \cite{valles}. For $f=1/3$, a
resistive transition is found at nonzero temperature, as expected
from early results of equilibrium MC simulations \cite{shih85}. The
estimated critical temperature is slightly below the melting
transition of the commensurate vortex lattice \cite{eg12},
suggesting two separated transitions. However, further work is
required to obtain a more accurate estimate and to investigate the
critical behavior in detail. For $f=1/2$, the results are consistent
with a transition scenario where the critical temperature vanishes
and the linear resistivity shows thermally activated behavior. The
thermal critical exponent $\nu$ of the zero-temperature transition
estimated from a dynamical scaling analysis is in fair agreement
with recent calculations from finite-size correlation length scaling
\cite{eg12}. A dynamical freezing at a lower temperature $T_f$  was
identified from deviations of the fluctuation-dissipation relation
between linear resistivity and voltage autocorrelations. It should
be pointed out that, since equilibrium data could not be obtained
below $T_f$, a resistive transition at much lower temperatures can
not be ruled out. Moreover, since the scaling analysis assumes a
second-order phase transition, a first-order resistive transition
near or below $T_f$ is also not excluded. The resistive behavior is
qualitatively consistent with experimental observations in
ultra-thin superconducting films with a triangular lattice of
nanoholes \cite{valles}, taking into account the effects of weak
Josephson-coupling disorder. A more quantitative comparison to the
experimental system, including geometrical disorder
\cite{gk86,benz88,korsh96,gupta99},  and the relation between the
resistive behavior and the vortex structure \cite{korshu} for
$f=1/2$, as well as $f=1/3$,  require further work.

\section{Acknowledgements}

This work was supported by Funda\c c\~ao de Amparo \`a Pesquisa do
Estado de S\~ao Paulo (FAPESP; grant No. 07/08492-9) and computer
facilities from Centro Nacional de Processamento de Alto Desempenho
em S\~ao Paulo (CENAPAD-SP).

%\begin{figure}
%\includegraphics[bb= 1cm -0cm  9cm   6.5cm,
%width=7.5cm]{scal_Jnl.eps} \caption{ Finite-size scaling plot of the
%crossover current $J_{nl}$ for different temperatures $T$ and
%systems sizes $L$, with $\nu =1.25$. } \label{scaljnl}
%\end{figure}

\end{document}